\documentclass[aps,prd,showpacs,onecolumn,preprintnumbers,notitlepage,
nofootinbib,amsfont,amssymb,10pt,singlespacing] {revtex4-1}
\usepackage{amsmath,bm}
\usepackage{times}
\usepackage{color,graphicx}
\usepackage{slashed}
\usepackage{hyperref}

\newcommand{\beq}{\begin{eqnarray}}
\newcommand{\eeq}{\end{eqnarray}}
\newcommand{\non}{\nonumber\\ }

\newcommand{\nsl}{ n \hspace{-1.8truemm}/ }
\newcommand{\lsl}{ l \hspace{-1.8truemm}/ }
\newcommand{\ksl}{ k \hspace{-1.9truemm}/ }

\def\lsim{ {\ \lower-1.2pt\vbox{\hbox{\rlap{$<$}\lower6pt\vbox{\hbox{$\sim$}
}}}\ } }
\def\gsim{ {\ \lower-1.2pt\vbox{\hbox{\rlap{$>$}\lower6pt\vbox{\hbox{$\sim$}
}}}\ } }

\definecolor{Red}{rgb}{1.,0.,0.}

\definecolor{Blue}{rgb}{0.,0.,1.}

\newcommand{\orcid}[1]{\thanks{\href{http://orcid.org/#1}{ORCID: #1}}}

\definecolor{nicered}{rgb}{0.7,0.1,0.2}
\definecolor{nicegreen}{rgb}{0.1,0.4,0.2}

\hypersetup{colorlinks,citecolor=nicegreen,linkcolor=nicered}

\begin{document}
\title{\boldmath Next-to-leading-logarithm $k_T$ resummation for $B_c\to J/\psi$ decays}
\author{Xin~Liu}
\email
[Electronic address: ]
{liuxin@jsnu.edu.cn}
\orcid{0000-0001-9419-7462}
\affiliation{Department of Physics,
Jiangsu Normal University, Xuzhou 221116, People's Republic of China}

\author{Hsiang-nan~Li}
\email
[Corresponding author: ]
{hnli@phys.sinica.edu.tw}
\affiliation{ Institute of Physics, Academia Sinica, Taipei, Taiwan 115, Republic of China}

\author{Zhen-Jun~Xiao}
\email
[Electronic address: ]
{xiaozhenjun@njnu.edu.cn}
\orcid{0000-0002-4879-209X}
\affiliation{ Department of Physics and Institute of Theoretical
Physics,\\Nanjing Normal University, Nanjing 210023, People's Republic of China}


\date{\today}

\begin{abstract}

We derive the $k_T$ resummation for a transverse-momentum-dependent charmonium wave function,
which involves the bottom quark mass $m_b$, the charm quark mass $m_c$, and the charm quark
transverse momentum $k_T$, up to the next-to-leading-logarithm (NLL) accuracy under the hierarchy
$m_b\gg m_c \gg k_T$. The resultant Sudakov factor is employed in the perutrbative
QCD (PQCD) approach to the $B_c\to J/\psi$ transition form factor $A_0^{B_c \to J/\psi}(0)$
and the $B_c^+ \to J/\psi \pi^+$ decay. We compare the NLL resummation effect on these processes
with the leading-logarithm one in the literature,  and find  a  $(5-10)\%$  enhancement to  the form factor
$A_0^{B_c \to J/\psi}(0)$ and a $(10-20)\%$  enhancement to the decay rate $BR(B_c^+ \to J/\psi \pi^+)$.
The improved $k_T$ resummation formalism is applicable to the PQCD analysis of heavy meson decays to other
charmonia.

\end{abstract}


\pacs{13.25.Hw, 12.38.Bx, 14.40.Nd}
\maketitle

\section{INTRODUCTION}

$B_c$ meson decays contain rich heavy quark dynamics in both perturbative
and nonperturbative regimes, that is worth of thorough exploration
with high precision. It is thus crucial to develop an appropriate
theoretical framework for analyzing $B_c$ meson decays, for which data have been
accumulated rapidly. A framework
available in the literature is the perturbative QCD (PQCD) approach, which
basically follows the conventional one for $B$ meson decays: the dependence on
the finite charm quark mass is included in hard decay kernels
but neglected in the $k_T$ resummation formula for meson wave functions \cite{Xiao:2013lia,
Sun:2014ika,Wang:2014yia,Rui:2014tpa,Rui:2015iia,Rui:2016opu,Liu:2017cwl, Sun:2017lla,Ma:2017aie,Rui:2017pre}.
Hence, a theoretical challenge from these decays is to derive the $k_T$ resummation
associated with energetic charm quarks with a finite mass. Such a rigorous $k_T$ resummation
formalism for a typical transition $B_c\to J/\psi$ was first attempted in \cite{Liu:2018kuo}.
The derivation relies on the power counting for the
involved multiple scales, the bottom (charm) quark mass $m_b$ ($m_c$)
and the parton transverse momentum $k_T$. We have adopted the power counting
rule proposed in \cite{Kurimoto:2002sb}, which obeys the hierarchy
$m_b\gg m_c\gg k_T$. An intermediate impact of this hierarchy
is that the large infrared logarithms $\ln(m_c/k_T)$,
in addition to the ordinary ones $\ln(m_b/k_T)$, appear in the
PQCD evaluation of $B_c$ meson decays, and need to be resummed.

To proceed with the $k_T$ resummation, we considered the $B_c\to J/\psi$ transition process,
constructed the transverse-momentum-dependent (TMD) $B_c$ and $J/\psi$ meson wave functions in
the $k_T$ factorization theorem \cite{Nagashima:2002ia,Li:2004ja},
and then performed the one-loop evaluation according to the wave-function
definition as a nonlocal hadronic matrix element.
The large logarithms attributed to the overlap of the collinear and
soft radiative corrections were found to differ from those in $B$ meson
decays into light mesons \cite{Li:1994cka}, because of the additional charm quark scale.
However, only the leading double logarithms
from the correction to the quark-Wilson-line vertex in meson wave functions were captured in
\cite{Liu:2018kuo}, namely, the $k_T$ resummation for the $B_c\to J/\psi$ decays was achieved
at the leading-logarithm (LL) accuracy so far. How the charm quark mass dependence in the LL
$k_T$ resummation affects the $B_c\to J/\psi$ transition form factor and the
$B_c^+ \to J/\psi \pi^+$ branching ratio was then investigated \cite{Liu:2018kuo}.

In this letter we will complete the $k_T$ resummation for the $B_c$ and $J/\psi$
meson wave functions up to the next-to-leading-logarithm (NLL) accuracy. Since the
analysis involves the convolution with the corresponding hard decay kernel at the NLL
accuracy, it is more convenient to perform the resummation in the impact parameter $b$ space,
which is conjugate to the transverse momentum $k_T$. We start with
the one-loop calculation for the $J/\psi$ meson wave function, from which all
important logarithms are identified. It is found that these logarithms are
grouped into two sets, $\ln(m_b b)$ and $\ln(m_c b)$, with their coefficients
being identical but opposite in sign. It hints that the resummation technique
can be applied to these two sets of logarithms separately: the $k_T$ resummation
is constructed for the first set, that for the second set
can be inferred trivially via the replacement of the argument $m_b$ by $m_c$,
and the final result is given by the difference between them.
Moreover, the resummation technique applied to the first set of logarithms
is the same as the one applied to a light meson case \cite{Botts:1989kf} without the
intermediate scale $m_c$. We emphasize that the matching to the one-loop $J/\psi$
meson wave function is crucial for achieving the NLL accuracy. The NLL $k_T$ resummation
for the $B_c$ meson wave function is then done in a similar way. At last, the NLL resummation
effect is employed in the PQCD evaluation of the $B_c \to J/\psi$ transition
form factor and the $B_c^+ \to J/\psi \pi^+$ branching ratio, and compared with
the LL effect observed in \cite{Liu:2018kuo}.

\section{$B_c$ AND $J/\psi$ MESON WAVE FUNCTIONS}

In this section we construct the definitions of the TMD wave functions for the $B_c$ and $J/\psi$ mesons.
Consider the $B_c(P_1)\to J/\psi(P_2)$ transition at the maximal recoil,
where
\begin{eqnarray}
P_1 = \frac{m_{B_c}}{\sqrt{2}} (1, 1, {\bf 0}_T)\;, \qquad
P_2 = \frac{m_{B_c}}{\sqrt{2}} (1, r_{J/\psi}^2, {\bf 0}_T), \label{bj}
\end{eqnarray}
label the $B_c$ and $J/\psi$ meson momenta in the light-cone coordinates,
respectively, with $r_{J/\psi} = m_{J/\psi}/m_{B_c}$,
$m_{B_c}$ ($m_{J/\psi}$) being the $B_c$ ($J/\psi$) meson mass.
Denote the anti-charm quark momenta as $k_1=x_1P_1$ in the $B_c$ meson,
and $k_2 = x_2P_2$ in the $J/\psi$ meson with the momentum fractions $x_1$ and $x_2$.
Allowing the charm quarks to be off-shell only slightly, i.e.,
$k_1^2-m_c^2\sim k_2^2-m_c^2 \sim O(m_c \Lambda_{\rm QCD})$,
$\Lambda_{\rm QCD}$ being the QCD scale, we have postulated that the shapes of the $B_c$
and $J/\psi$ meson wave functions exhibit peaks
around $x_1\sim m_c/m_b$ and $x_2\sim 1/2$, respectively \cite{Liu:2018kuo}.
A charm quark, carrying a longitudinal momentum initially, gains transverse momenta
by radiating gluons \cite{Nagashima:2002ia}, which generate the $k_T$
dependence of a TMD wave function.

We point out that the power counting for a parton transverse momentum $k_T$ is nontrivial,
compared to the power counting for the fixed mass
scales like $m_b$ and $m_c$. First, the $k_T$ factorization is suitable for a multi-scale
process, such as the region of a small parton momentum fraction $x$
in a process with a large scale $Q$. The small $x$ introduces an additional intermediate scale
$xQ^2\sim Q\Lambda_{\rm QCD}$, respecting the hierarchy $Q^2\gg xQ^2\gg\Lambda_{\rm QCD}^2$.
A parton $k_T$, being an integration variable in a $k_T$ factorization formula, can take values of orders
of the above scales. The criteria for applying the $k_T$ factorization include: 1) the hard kernel of a
considered process depends on the large scale $Q^2$ and the intermediate scale $Q\Lambda_{\rm QCD}$,
but not on the small scale $\Lambda_{\rm QCD}^2$; 2) the factorization of the relevant
TMD wave functions hold for a parton $k_T$ at both the intermediate and small
scales. Once these two criteria are satisfied, the $k_T$ dependence in the hard kernel is not
negligible, and a convolution between the hard kernel and the TMD wave functions can be derived.
If the hard kernel involves only the large scale, the $k_T$ dependence of the hard kernel can be
neglected. It is then integrated out in the wave functions, and one is led to the collinear
factorization. An integrated TMD wave function can be written as a convolution of its corresponding
distribution amplitude with a perturbative kernel \cite{Collins:2011zzd}.
A typical example for establishing the $k_T$ factorization at the one-loop level
is referred to \cite{Nandi:2007qx}: the matching between the QCD diagrams and the pion TMD
wave function for the $\pi\gamma^*\to\gamma$ process gives the hard kernel in Eq.~(40) of
\cite{Nandi:2007qx}, that depends on $Q^2$ and $xQ^2+k_T^2$, but not on
$k_T^2$. Namely, the resultant hard kernel satisfies the first criterion, no matter which scale of $k_T$ is.
It can be shown that the eikonalization for factorizing the pion TMD wave function in Eq.~(21) of
\cite{Nandi:2007qx} holds, as required by the second criterion, for both
$k_T^2\sim l_T^2\sim O(Q\Lambda_{\rm QCD})$ and $O(\Lambda_{\rm QCD}^2)$,
which are lower than the dominant invariant $Q^2$ in that equation.


Since a TMD wave function contains the contributions characterized by both the intermediate and small
scales, it is legitimate to further factorize the former out of the wave function, as the intermediate scale
is regarded as being perturbative. Motivated by this observation, the joint resummation which organizes
the mixed logarithms formed by the two invariants $xQ^2$ and $k_T^2$ has been performed for the
pion wave function \cite{Li:2013xna}. This resummation, like an ordinary $k_T$
resummation, is justified perturbatively for the scale $k_T^2\sim O(Q\Lambda_{\rm QCD})$.
After this organization, the remaining pion wave function involves only the small scale $\Lambda_{\rm QCD}^2$.
Similarly, it is also legitimate to further factorize the contribution characterized by
an intermediate scale out of a hard kernel in the $k_T$ factorization. This re-factorization
yields the jet function defined in \cite{Li:2001ay}, through which the logarithms of $xQ^2$
are resummed to all orders.

Because more scales are involved in the $B_c\to J/\psi$ transition than in the $\pi\gamma^*\to\gamma$
process, there exist more leading infrared regions than in the latter. It has been argued
\cite{Liu:2018kuo} that a collinear region for the $B_c\to J/\psi$
transition is described by the power counting
\begin{eqnarray}
l^\mu=(l^+,l^-,l_T)\sim \left(\frac{m_b}{m_c}\Lambda, \frac{m_c}{m_b}\Lambda,\Lambda\right),
\label{pow}
\end{eqnarray}
where $l$ is the momentum of a radiative gluon with the invariant mass squared
of $O(\Lambda^2)$, and $\Lambda=m_c$ or $k_T$ represents a lower scale.
If the relation $m_bk_T\sim m_c^2$ is assumed \cite{Wang:2017jow},
a collinear gluon momentum $l^\mu\sim (m_b k_T/m_c, m_ck_T/m_b,k_T)$
for $\Lambda=k_T$ will be equivalent to $l^\mu\sim (m_c, k_T^2/m_c,k_T)$.
The soft radiation in the $B_c\to J/\psi$ transition is
characterized by $l^\mu\sim (\Lambda, \Lambda,\Lambda)$ \cite{Liu:2018kuo}.
The dominant collinear (soft) enhancement is absorbed into the $J/\psi$ ($B_c$) meson wave
function, and the remaining contribution goes into a hard decay kernel.
It is easy to see that the hard kernel for the $B_c\to J/\psi$ transition
involves the scales down to $m_c^2$: the hard gluon invariant mass squared
$(k_1-k_2)^2$ contains $k_1^2\sim k_2^2 \sim m_c^2$. It will be explained that the
$B_c$ and $J/\psi$ meson wave functions can be factorized for the scales up to $m_c^2$.
The above observations indicate that the $k_T$ factorization is appropriate for the analysis of
the $B_c\to J/\psi$ transition. As the scale $m_c$ is regarded
as perturbative \cite{Wang:2017jow}, a hard piece with the scale $m_c$
can be further factorized out of the wave-function definitions. Such a re-factorization has been applied
to the light-cone distribution amplitudes of doubly-heavy mesons, which are then expressed as
products of perturbatively calculable distribution parts and non-relativistic QCD (NRQCD) matrix
elements \cite{Xu:2016dgp,Wang:2017bgv}.

\begin{figure}[!!htb]
\centering
\begin{tabular}{l}
\includegraphics[width=0.8\textwidth]{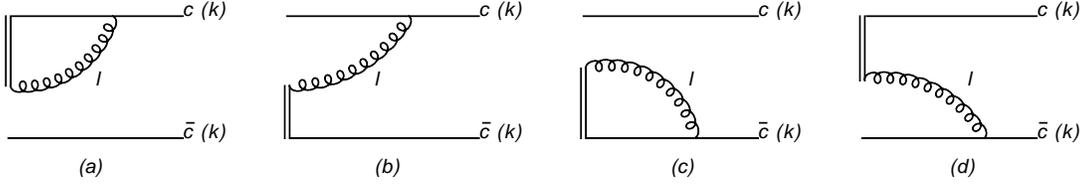}
\end{tabular}
\caption{One-loop vertex corrections to the $J/\psi$ meson wave function.}
  \label{fig:fig1}
\end{figure}

The collinear gluons in the $B_c\to J/\psi$ transition can be collected by a gauge link,
which is required by the gauge invariance
of a TMD wave function. The one-loop diagram, in which a radiative gluon attaches to
the valence $b$ quark in the $B_c$ meson and the valence $c$ quark in the $J/\psi$ meson,
gives the collinear enhancement from Eq.~(\ref{pow}).
The $b$ quark line is eikonalized in this region into $n^\nu/n\cdot l$, with $\nu$ labelling
the gluon vertex, and a gauge link in the direction $n=(1,1,{\bf 0}_T)/\sqrt{2}$
along the $B_c$ meson momentum is generated as shown in
Fig.~\ref{fig:fig1}(a). The diagram, in which a radiative
gluon attaches to the $\bar c$ quark in the $B_c$ meson and the $c$ quark in the $J/\psi$ meson,
also contains the collinear enhancement from Eq.~(\ref{pow}).
The eikonalization of the $\bar c$ quark line
in the $B_c$ meson results in a gauge link in the direction $n$ shown in Fig.~\ref{fig:fig1}(b).
The appearance of the gauge links in
Figs.~\ref{fig:fig1}(c) and \ref{fig:fig1}(d) can be explained similarly.
The Fierz identity is then inserted to factorize the fermion flow \cite{Nagashima:2002ia}
between the one-loop effective diagrams in Fig.~\ref{fig:fig1}
and the other parts of the transition process .
Note that the above factorization procedure holds for $k_T^2$ up to the scale $m_c^2$,
which is lower than the invariant mass squared of $O(m_b^2)$ ($O(m_bm_c)$) of
the eikonalized $b$ ($\bar c$) quark.

The $J/\psi$ meson wave function $\Phi_{J/\psi}$ factorized out of the above transition process
depends on two external vectors, the momentum $P_2$ and the direction $n$ of the gauge link.
Since the Feynman rule for the gauge link is scale-invariant in $n$, $\Phi_{J/\psi}$
must depend on $n$ through the ratio of the Lorentz invariants, $\xi^2=4(P_2\cdot n)^2/|n^2|$
\cite{Li:1996gi,Li:2013ela}.
We define the $J/\psi$ meson wave function $\Phi_{J/\psi}(x,k_T,\xi^2,m_c)$ as
\begin{eqnarray}
\Phi_{J/\psi}(x,k_T,\xi^2,m_c)&=&\int\frac{d^4y}{(2\pi)^3}
e^{-ik\cdot y}\langle 0|{\bar c}(y) W_y(n)^{\dag}
\nsl_{-}W_0(n)c(0)|J/\psi(P_2)\rangle\delta(u\cdot y),
\label{de1}
\end{eqnarray}
where $k=(xP_2^+,xP_2^-,{\bf k}_T)$ is the $\bar c$ quark momentum, and $y=(y^+,y^-,{\bf y}_T)$
denotes the coordinate of the $\bar c$ quark field. The projector $\nsl_{-}$ with the null
vector $n_-=(0,1,{\bf 0}_T)$ along the minus direction arises from the aforementioned insertion of the
Fierz identity. The integration over the momentum orthogonal to the $\bar c$ quark momentum leads to
the function $\delta(u\cdot y)$, with dimensionless vector
$u=(-1/r_{J/\psi}^2,1,{\bf 0}_T)/\sqrt{2}$, which specifies the location of $\bar c$ on the $y^+$-$y^-$ plane.
Besides, $\Phi_{J/\psi}$ also depends on the factorization scale $\mu_{f}$, which is not shown explicitly.
The factor $W_y(n)$ represents the gauge link operator
\begin{eqnarray}
\label{eq:WL.def} W_y(n) = P \exp\left[-ig \int_0^\infty d\lambda
n\cdot A(y+\lambda n)\right].
\end{eqnarray}
A vertical link to connect the two links $W_y(n)$ and $W_0(n)$ at infinity is implicit.
The removal of the gauge-link self-energy corrections \cite{Li:2014xda} from the definition in Eq.~(\ref{de1})
is understood, which is, however, irrelevant to the $k_T$ resummation to be performed in the
next section.

The soft region characterized by the power counting
$l^\mu\sim (\Lambda, \Lambda,\Lambda)$ with $\Lambda$ denoting $m_c$ or $k_T$, also contributes
dominantly to the one-loop diagrams discussed above. This contribution is
factorized into the $B_c$ meson wave function by eikonalizing the charm
quark line on the $J/\psi$ meson side, and the same gauge link in the direction $n$
is chosen. Under the considered hierarchy $m_b\gg m_c$, the $B_c$ meson
wave function can be defined in a way similar to the $B$ meson wave function,
\begin{eqnarray}
\Phi_{B_c}(x,k_T,\xi^2,m_c)&=&\int\frac{d^4y}{(2\pi)^3}
e^{-ik\cdot y}\langle 0|{\bar c}(y) W_y(n)^{\dag}\gamma_5\nsl_{+} W_0(n)
b(0)|B_c(P_1)\rangle\delta(u'\cdot y),
\label{bc}
\end{eqnarray}
where $k=(xP_1^+,xP_1^-,{\bf k}_T)$ is the $\bar c$ quark momentum, and the projector
$\gamma_5\nsl_{+}$ arises from the insertion of the Fierz identity.
The dimensionless vector $u'=(-1,1,{\bf 0}_T)/\sqrt{2}$, i.e., a vector in the $z$ direction,
is introduced to specify the location of $\bar c$ on the $y^+$-$y^-$ plane: ${\bar c}$ is located
on the time axis in this case.
If one adopts an alternative power counting for the involved heavy quark masses, $m_b\sim m_c$,
it will be more appropriate to define the $B_c$ meson wave function in
the effective theory of NRQCD rather than in QCD directly, since
the separation between the bottom and charm quarks in coordinate space is much larger
than $1/m_b$.

To identify the important logarithms in the $J/\psi$ meson wave function, we calculate the
one-loop effective diagrams displayed in Fig.~\ref{fig:fig1} with an on-shell charm quark.
Though the factorization of the $J/\psi$ wave functions holds for $k_T$ up to the scale $m_c$,
we consider the hierarchy $m_b\gg m_c\gg k_T$, which generates the largest logarithms.
Assume that the $\bar c$ quark carries the momentum $k=xP_2$, and the $c$
quark carries $\bar k\equiv P_2-k=(1-x)P_2$. Figure~\ref{fig:fig1}(a), which does not induce a
transverse momentum of the charm quark, gives the loop integral
\beq
\Phi_a^{(1)}  &=& - \frac{i}{4} g^2 \mu_f^{2\epsilon}
\int \frac{d^{4-2\epsilon}l}{(2\pi)^{4-2\epsilon}}
{\rm tr}\biggl[ \nsl_{+} \frac{\bar \ksl + \lsl + m_c }{(\bar k +l)^2-m_c^2}
\gamma_\nu \nsl_{-}  \biggr] \frac{1}{l^2-m_g^2} \frac{n^\nu}{n\cdot l},\label{phid}
\eeq
with the color factor $C_F=4/3$, the on-shell condition $\bar k^2\approx m_c^2$,
the factorization scale $\mu_f$, the gluon
momentum $l$, and the gluon mass $m_g$ as an infrared regulator. The projectors
$\nsl_{+}$ and $\nsl_{-}$ select the leading twist contribution.
A straightforward computation yields
\beq
\Phi_a^{(1)}
&=&\frac{\alpha_s}{4\pi}C_F \left[\frac{1}{\epsilon} + \ln \frac{4\pi \mu_f^2}{m_c^2 e^{\gamma_E}}
-2\ln\frac{(1-x)^2\xi^2}{m_c^2}\ln\frac{(1-x)^2\xi^2}{m_g^2}
+\ln\frac{(1-x)^2\xi^2}{m_c^2}+ 2 - \frac{\pi^2}{3} \right],
\label{phid1}
\eeq
where $1/\epsilon$ represents an ultraviolet divergence and $\gamma_E$ is the Euler constant.
It is found that the collinear divergence
regularized by the charm quark mass $m_c$ and the soft divergence regularized
by the gluon mass $m_g$ overlap to produce the product of the corresponding logarithms in the
above expression.

For Fig.~\ref{fig:fig1}(b), the transverse loop momentum $l_T$, flowing through the hard
decay kernel, is not negligible in the $k_T$ factorization
as explained before. To facilitate the loop calculation, we apply the
Fourier transformation to turn the convolution between the hard kernel and the
$J/\psi$ meson wave function into a product, and write the integral for the latter in
the impact parameter $b$ space as
\beq
\Phi_b^{(1)}  = \frac{i}{4} g^2 C_F
\int \frac{d^{4}l}{(2\pi)^{4}}\exp(i{\bf l}_T\cdot {\bf b})
{\rm tr}\biggl[ \nsl_{+} \frac{\bar \ksl + \lsl + m_c }{(\bar k +l)^2-m_c^2}
\gamma_\nu \nsl_{-}  \biggr] \frac{1}{l^2-m_g^2} \frac{n^\nu}{n\cdot l}\;. \label{phie}
\eeq
After performing the integration, we get
\beq
\Phi_b^{(1)}  =\frac{\alpha_s}{4\pi}C_F \left[\frac{1}{2}\ln^2\frac{(1-x)^2\xi^2}{m_c^2}
+2\ln\frac{(1-x)^2\xi^2}{m_c^2}\ln\frac{2(1-x)\xi}{b m_g^2 e^{\gamma_E}}\right].
\label{phie1}
\eeq
Compared to Eq.~(\ref{phid1}), the above expression is free of an ultraviolet divergence
due to the Fourier factor $\exp(i{\bf l}_T\cdot {\bf b})$.
Note that the integration over the transverse momentum $l_T$ in the presence of
$\exp(i{\bf l}_T\cdot {\bf b})$ generates a Bessel function $K_0$, which can be
approximated by a logarithmic function as its argument approaches to zero. Hence,
Eq.~(\ref{phie1}) is valid only up to the logarithmic term, strictly speaking. This
approximation works well enough for the matching between the NLL resummation and the
one-loop result.

The sum of Eqs.~(\ref{phid1}) and (\ref{phie1}) gives
\beq
\Phi_{a+b}^{(1)} &=&\frac{\alpha_s}{4\pi}C_F \left[\frac{1}{\epsilon}
 + \ln \frac{4\pi \mu_f^2}{m_c^2 e^{\gamma_E}}+\frac{1}{2}\ln^2\frac{(1-x)^2\xi^2}{m_c^2}
-2\ln\frac{(1-x)^2\xi^2}{m_c^2}\ln\frac{(1-x)\xi b e^{\gamma_E}}{2}
+\ln\frac{(1-x)^2\xi^2}{m_c^2}+ 2 - \frac{\pi^2}{3}\right]\non
&=&\frac{\alpha_s}{4\pi}C_F \left[\frac{1}{\epsilon}
 + \ln \frac{4\pi \mu_f^2}{m_c^2 e^{\gamma_E}}
 -\frac{1}{2}\ln^2\frac{(1-x)^2\xi^2b^2e^{2\gamma_E-1}}{4}
 +\frac{1}{2}\ln^2\frac{m_c^2b^2e^{2\gamma_E-1}}{4}
+ 2 - \frac{\pi^2}{3}\right].
\label{ab}
\eeq
It is seen in the first line that the infrared regulator $m_g$ has been cancelled
as expected, and the soft scale has been replaced by $1/b$. It implies that the color
transparency argument holds, and the soft divergences disappear in the summation over diagrams.
The ultraviolet logarithm can be removed
by choosing the factorization scale $\mu_f=m_c$, which defines the initial scale for the evolution of
the $J/\psi$ meson wave function in $\mu_f$. In the second line we have reorganized the sum
into the desired form: the logarithms are grouped into two sets,
one containing $\ln^2(\xi b)$ and another containing $\ln^2(m_c b)$, as postulated in
\cite{Liu:2018kuo}. The difference between them arises only from the arguments
$(1-x)\xi$ and $m_c$, and from the sign. This must be the case, because
the collinear and soft divergences are regularized by the charm mass $m_c$ and the impact parameter
$1/b$, respectively, in the present calculation. Hence, the overlap of the corresponding
logarithms should not generate $\ln^2 (1/b)$, such that the above two sets of double logarithms
have equal coefficients but with opposite signs.
The sum of the contributions from Figs.~\ref{fig:fig1}(c) and \ref{fig:fig1}(d)
can be obtained simply by substituting the momentum fraction $x$ for $(1-x)$ in
Eq.~(\ref{ab}),
\beq
\Phi_{c+d}^{(1)} &=&\frac{\alpha_s}{4\pi}C_F \left[\frac{1}{\epsilon}
 + \ln \frac{4\pi \mu_f^2}{m_c^2 e^{\gamma_E}}
 -\frac{1}{2}\ln^2\frac{x^2\xi^2b^2e^{2\gamma_E-1}}{4}
 +\frac{1}{2}\ln^2\frac{m_c^2b^2e^{2\gamma_E-1}}{4}
+ 2 - \frac{\pi^2}{3}\right].\label{cd}
\eeq

A remark is in order. It has been elaborated recently \cite{Forte:2020fbc} that the coefficient of a
double logarithm associated with an on-shell parton is half of the coefficient
in the off-shell case. Taking Fig.~\ref{fig:fig1}(a) as an example, we have
evaluated its contribution for an energetic charm quark off-shell by $-k_T^2$,
and obtained \cite{Liu:2018kuo}
\beq
\Phi_a^{(1)}&=& \frac{\alpha_s}{4\pi} C_F
\left[ \frac{1}{\epsilon} + \ln \frac{4\pi \mu_f^2}{m_c^2 e^{\gamma_E}}
 -\ln^2{\frac{(1-x)^2\xi^2}{k_T^2}}+ \ln^2{\frac{m_c^2}{k_T^2}}
 + \ln{\frac{(1-x)^2\xi^2}{m_c^2}} + 2 - \frac{2}{3}\pi^2  \right].
\label{eq:phid}
\eeq
Comparing Eq.~(\ref{eq:phid}) with Eq.~(\ref{phid1}), we indeed find that the coefficients
of the double logarithms have been reduced to half in the on-shell case.
We explain that the fixed-order calculation with an off-shell quark is required for
the proof of the $k_T$ factorization \cite{Nagashima:2002ia}, in which the common parton virtuality
$-k_T^2$ is adopted to regularize the infrared divergences in both QCD and
effective diagrams. The $k_T$ factorization holds, if the infrared logarithms
$\ln k_T^2$ could be shown to cancel between these two sets of diagrams \cite{Nagashima:2002ia}.
As deriving the $k_T$ resummation formula, we consider an on-shell initial parton, which
becomes virtual by transverse momenta through radiations. The resummation
technique aims at collecting these radiations to all orders.

\section{NLL $k_T$ RESUMMATION}

\begin{figure}[!!htb]
\centering
\begin{tabular}{l}
\includegraphics[width=0.6\textwidth]{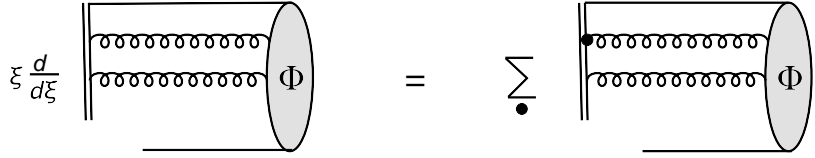}
\end{tabular}
\caption{Graphic representation of the derivative $\xi d\Phi_{J/\psi}/d\xi$.}
  \label{fig:fig2}
\end{figure}

We first proceed with the NLL $k_T$ resummation for the $J/\psi$ meson wave function
based on the complete one-loop results in the impact parameter space presented in the previous section,
assuming $\xi\gg m_c \gg 1/b\gg \Lambda_{\rm QCD}$.
The strategy is to focus only on the first set of logarithms $\ln(\xi b)$,
whose treatment is similar to that of a light meson case, and then infer the
resummation formula for the second set via the replacement of $\xi$ by $m_c$.
The choice of $n$ is arbitrary in principle, which does not affect the
collection of the collinear enhancement. This is the key observation for performing the resummation.
We then study the variation of the $J/\psi$ meson wave function with the gauge link direction $n$,
which is equivalent to the variation with the dominant component $P_2^+$
of the $J/\psi$ meson momentum via the scale $\xi$,
\begin{equation}
P_2^+\frac{d}{dP_2^+}\Phi_{J/\psi}=
\xi\frac{d}{d\xi}\Phi_{J/\psi}=-\frac{n^2}{P_2\cdot n}P_2^\alpha\frac{d}{dn^\alpha}\Phi_{J/\psi}.
\label{cr}
\end{equation}
The technique of varying gauge links has been applied to the resummation of various
types of logarithms, such as the rapidity logaritms in the $B$ meson wave function \cite{Li:2012md},
and the joint logarithms in the pion wave function \cite{Li:2013xna}.
The differentiation of each eikonal vertex and of its associated eikonal propagator on the gauge link
with respect to $n_\alpha$,
\begin{eqnarray}
-\frac{n^2}{P_2\cdot n}P_2^\alpha\frac{d}{dn^\alpha}\frac{n^\mu}{n\cdot l}
=\frac{n^2}{p\cdot n}\left(\frac{P_2\cdot l}{n\cdot l}n^\mu-P_2^\mu\right)
\frac{1}{n\cdot l}
\equiv\frac{{\hat n}^\mu}{n\cdot l}\;,
\label{dp}
\end{eqnarray}
leads to the derivative $\xi d\Phi_{J/\psi}/d\xi$ depicted in Fig~\ref{fig:fig2}. The summation in
Fig.~\ref{fig:fig2} includes different attachments of the new vertex ${\hat n}^\mu$
defined by the last expression in Eq.~(\ref{dp}), and represented by the symbol ``$\bullet$".

\begin{figure}[!!htb]
\centering
\begin{tabular}{l}
\includegraphics[width=0.8\textwidth]{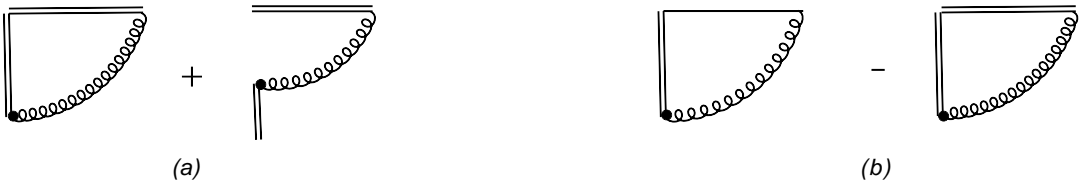}
\end{tabular}
\caption{(a) soft function and (b) hard function at $O(\alpha_s)$.}
  \label{fig:fig3}
\end{figure}

As stated before, terms of $O(m_c^2)$ can be dropped for the resummation of the first
set of logarithms. If the loop momentum $l$ is parallel to $P_2$, the factor $P_2\cdot l$, being
of $O(m_c^2)$, is negligible. When the second term $P_2^\mu$ in ${\hat n}^\mu$ is
contracted with a vertex in $\Phi_{J/\psi}$, where all momenta are mainly parallel
to $P_2$, the contribution from this collinear region is also of
$O(m_c^2)$, and negligible. That is, the leading regions of $l$ are soft and hard.
According to \cite{Li:1996gi}, as the loop momentum flowing through the new vertex
is soft, only the diagram with the new vertex being located at the outer most end
of the gauge link dominates, and gives large single logarithms. In this soft
region the subdiagram containing the new vertex can be factorized using the eikonal approximation,
and the remainder is assigned to $\Phi_{J/\psi}$.
This subdiagram is absorbed into a soft function $K$, whose $O(\alpha_s)$ contribution
is displayed in Fig.~\ref{fig:fig3}(a). As the loop momentum flowing through
the new vertex is hard, only the diagram with the new vertex being located at the
inner most end of the gauge link dominates. In this region the subdiagram
containing the new vertex is factorized
into a hard function $G$, whose $O(\alpha_s)$ contribution is displayed
in Fig.~\ref{fig:fig3}(b), and the remainder
is identified to be $\Phi_{J/\psi}$.

We arrive at the differential equation in the impact parameter space
\begin{equation}
P_2^+\frac{d}{dP_2^+}\Phi_{J/\psi}=2\left[K(b\mu,\alpha_s(\mu))+G
(P_2^+/\mu,\alpha_s(\mu))\right]\Phi_{J/\psi},
\label{dph}
\end{equation}
where the arguments of $K$ and $G$ specify their characteristic scales.
Figure~\ref{fig:fig3}(a) contributes
\begin{eqnarray}
K&=&-ig^2C_F\mu^\epsilon\int\frac{d^{4-\epsilon} l}
{(2\pi)^{4-\epsilon}}\frac{{\hat n}^\mu}{n\cdot l}
\frac{g_{\mu\nu}}{l^2}\frac{P_2^\nu}{P_2\cdot l}[1-\exp(i{\bf l}_T\cdot {\bf b})]-\delta K,
\label{k1}
\end{eqnarray}
with $\delta K$ being an additive counterterm. The Fourier factor $\exp(i{\bf l}_T\cdot {\bf b})$
appears in the second diagram of Fig.~\ref{fig:fig3}(a), because the loop momentum flows through
$\Phi_{J/\Psi}$, such that the scale $1/b$ serves as an infrared cutoff of the loop integral
in Eq.~(\ref{k1}). The $O(\alpha_s)$ contribution to $G$ from Fig.~\ref{fig:fig3}(b),
where the soft subtraction is to avoid double counting of the soft contribution,
is written as
\begin{eqnarray}
G=-ig^2C_F\mu^\epsilon\int\frac{d^{4-\epsilon} l}
{(2\pi)^{4-\epsilon}}\frac{{\hat n}^\mu}{n\cdot l}\frac{g_{\mu\nu}}{l^2}
\left[\frac{\not P_2+\not l}{(P_2+l)^2}\gamma^\nu-\frac{P_2^\nu}{P_2\cdot l}\right]
-\delta G\;,
\label{g1}
\end{eqnarray}
where the charm quark mass $m_c$ has been dropped as explained before, and
$\delta G$ is an additive counterterm. Choosing the subtraction scheme
\begin{eqnarray}
\delta{K}=-\frac{\alpha_s}{2\pi}C_F\left[\frac{2}{\epsilon}
+\ln (\pi e^{\gamma_E})\right]=-\delta G,
\label{dkp1}
\end{eqnarray}
we get the soft and hard functions
\begin{eqnarray}
{K}&=&-\frac{\alpha_s}{2\pi}C_F\ln(b^2\mu^2),\nonumber\\
{G}&=&-\frac{\alpha_s}{2\pi}C_F\ln\frac{\xi^2e^{2\gamma_E-1}}{4\mu^2}.
\label{50}
\end{eqnarray}

Since $K$ and $G$ contain only single soft and ultraviolet logarithms,
respectively, they can be treated by RG methods:
\begin{equation}
\mu\frac{d}{d\mu}K=-\lambda_K=
-\mu\frac{d}{d\mu}G,
\label{kg}
\end{equation}
in which the anomalous dimension of $K$, $\lambda_K=\mu d\delta K/d\mu$,
is given, up to two loops, by \cite{KT82}
\begin{equation}
\lambda_K(\alpha_s)=\frac{\alpha_s}{\pi}C_F+\left(\frac{\alpha_s}{\pi}
\right)^2C_F\left[C_A\left(\frac{67}{36}
-\frac{\pi^{2}}{12}\right)-\frac{5}{18}n_{f}\right],
\label{lk}
\end{equation}
with the number of quark flavors $n_{f}$ and the color
factor $C_A=3$. As solving Eq.~(\ref{kg}), we allow the scale $\mu$ to evolve to
the infrared cutoff $1/b$ in $K$ and to $P_2^+$ in $G$, and obtain
the RG solution
\begin{eqnarray}
K(b\mu,\alpha_s(\mu))+G(P_2^+/\mu,\alpha_s(\mu))
&=&K(1,\alpha_s(1/b))+G(1,\alpha_s(P_2^+))-
\int_{1/b}^{P_2^+}\frac{d{\bar\mu}}{\bar\mu}
\lambda_K(\alpha_s({\bar\mu})),\nonumber \\
&=&-\frac{\alpha_s(P_2^+)}{2\pi}C_F\ln \frac{e^{2\gamma_E-1}}{2}-
\int_{1/b}^{P_2^+}\frac{d{\bar\mu}}{\bar\mu}\lambda_K(\alpha_s({\bar\mu})).
\label{skg}
\end{eqnarray}
The relation $\xi^2=2P_2^{+2}n^-/n^+=2P_2^{+2}$ for $n^+=n^-$ specified in Eq.~(\ref{de1})
has been inserted to get the initial condition $G(1,\alpha_s(P_2^+))$.

Substituting Eq.~(\ref{skg}) into Eq.~(\ref{dph}), we derive
\begin{eqnarray}
\Phi_{J/\psi} &=&
\exp\left[-\int_{1/b}^{(1-x)P_2^+}\frac{d \bar p}{\bar p}
\left(\int_{1/b}^{\bar p}\frac{d{\bar \mu}}{\bar \mu}
\lambda_{K}(\alpha_s({\bar \mu}))
+\frac{\alpha_s(\bar p)}{2\pi}C_F\ln \frac{e^{2\gamma_E-1}}{2}\right)\right]
\nonumber \\
& &\times \exp\left[-\int_{1/b}^{xP_2^+}\frac{d \bar p}{\bar p}
\left(\int_{1/b}^{\bar p}\frac{d{\bar \mu}}{\bar \mu}
\lambda_{K}(\alpha_s({\bar \mu}))
+\frac{\alpha_s(\bar p)}{2\pi}C_F\ln \frac{e^{2\gamma_E-1}}{2}\right)\right]\Phi_{J/\Psi}(x,b).
\label{sph}
\end{eqnarray}
We have set the lower bound of the variable $\bar p$ to $1/b$, and the upper
bounds to $(1-x)P_2^+$ and $xP_2^+$ for the integrals associated with Figs.~\ref{fig:fig1}(a)
and~\ref{fig:fig1}(b), and Figs.~\ref{fig:fig1}(c) and \ref{fig:fig1}(d), respectively,
so that the initial condition $\Phi_{J/\Psi}(x,b)$ depends on $x$ and $b$.
As pointed out before, the $k_T$ resummation formula for the second set of
important logarithms can be inferred from Eq.~(\ref{sph}) by substituting $m_c$ for
$(1-x)\xi$ and $x\xi$, namely, $m_c/\sqrt{2}$ for the upper
bounds of $\bar p$, and flipping the signs of the integrands. Combining the two resummation formulas,
we get the final result
\begin{eqnarray}
\Phi_{J/\psi}(x,b,\xi,m_c)&=&
\exp\left[-\int_{m_c/\sqrt{2}}^{(1-x)P_2^+}\frac{d \bar p}{\bar p}
\left(\int_{1/b}^{\bar p}\frac{d{\bar \mu}}{\bar \mu}
\lambda_{K}(\alpha_s({\bar \mu}))
+\frac{\alpha_s(\bar p)}{2\pi}C_F\ln \frac{e^{2\gamma_E-1}}{2}\right)\right]
\nonumber \\
& &\times \exp\left[-\int_{m_c/\sqrt{2}}^{xP_2^+}\frac{d \bar p}{\bar p}
\left(\int_{1/b}^{\bar p}\frac{d{\bar \mu}}{\bar \mu}
\lambda_{K}(\alpha_s({\bar \mu}))
+\frac{\alpha_s(\bar p)}{2\pi}C_F\ln \frac{e^{2\gamma_E-1}}{2}\right)\right]\Phi_{J/\psi}(x,b),
\label{sphm}
\end{eqnarray}
where the initial condition $\Phi_{J/\psi}(x,b)$ depends on the intermediate scale $m_c$
only via the factorization scale, i.e., the argument of the strong coupling $\alpha_s$ implicitly.
Expanding Eq.~(\ref{sphm}) to $O(\alpha_s)$ for a constant $\alpha_s$, we reproduce
all the logarithms in Eqs.~(\ref{ab}) and (\ref{cd}).
The remaining constant pieces will go into the $O(\alpha_s)$ hard decay kernel, when the one-loop
$J/\psi$ meson wave function and the one-loop decay amplitude are matched.

The above expression represents the complete NLL $k_T$ resummation for the $J/\psi$ meson wave
function, which involves the three scale $m_b$, $m_c$ and $k_T$. Compared to
\cite{Liu:2018kuo}, we have included the so-called $B$ term, ie.,
the second terms in the exponents in Eq.~(\ref{sphm}),
and determined the order-unity coefficient associated with the lower bound  of the variable $\bar p$ to
be $1/\sqrt{2}$, both of which correspond to NLL effects. The inclusion of these NLL pieces
requires a complete one-loop calculation of the $J/\psi$ meson wave function in the impact
parameter $b$ space. The $k_T$ resummation formula for the spectator charm quark in the $B_c$
meson then reads
\begin{eqnarray}
\Phi_{B_c}(x,b,\xi,m_c)&=&
\exp\left[-\int_{m_c/\sqrt{2}}^{xP_1^-}\frac{d \bar p}{\bar p}
\left(\int_{1/b}^{\bar p}\frac{d{\bar \mu}}{\bar \mu}
\lambda_{K}(\alpha_s({\bar \mu}))
+\frac{\alpha_s(\bar p)}{2\pi}C_F\ln \frac{e^{2\gamma_E-1}}{2}\right)\right]\Phi_{B_c}(x,b,m_c),
\label{bm}
\end{eqnarray}
according to the second line of Eq.~(\ref{sphm}),
for which the relevant large longitudinal component of the spectator momentum is $xP_1^-$.
Because the upper and lower bounds of the integration variable $\bar p$ are both of $O(m_c)$,
the resummation effect from Eq.~(\ref{bm}) is less significant.

\begin{table}[htb]
\caption{Dependence on the shape parameter $\beta_{B_c}$
of the quantities $A_0^{B_c \to J/\psi}(0)$ and
${\rm BR}(B_c^+ \to J/\psi \pi^+)$
in the PQCD approach at the LL and NLL accuracy. }
\label{tab:d-beta-A0-Br}
 \begin{center}\vspace{-0.5cm}{
\begin{tabular}[t]{c||c|c||c|c}
\hline  \hline
 Quantities   & \multicolumn{2}{c||} {$A_0^{B_c \to J/\psi}(0)$}
   & \multicolumn{2}{c}{${\rm BR}(B_c^+ \to J/\psi \pi^+)$} \\
\hline
  Shape parameter   &  LL  & NLL & LL & NLL
\\
\hline \hline
 $\beta_{B_c} =0.8$~ GeV
     &$0.488 - {\it i} 0.095$
     & $0.511 - {\it i} 0.147$
     &$2.80 \times 10^{-3}$
     & $3.10 \times 10^{-3}$
 \\
 $\beta_{B_c} = 0.9$~ GeV
     &$0.434 - {\it i} 0.070$
     & $0.460 - {\it i} 0.114$
     &$2.10 \times 10^{-3}$
     & $2.39 \times 10^{-3}$
 \\
 $\beta_{B_c} = 1.0$~ GeV
     &$0.384 - {\it i} 0.053$
     & $0.414 - {\it i} 0.090$
     &$1.60 \times 10^{-3}$
     & $1.87 \times 10^{-3}$
 \\
 $\beta_{B_c} = 1.1$~ GeV
     &$0.341 - {\it i} 0.039$
     & $0.373 - {\it i} 0.071$
     &$1.23 \times 10^{-3}$
     & $1.46 \times 10^{-3}$
 \\
 $\beta_{B_c} = 1.2$~ GeV
     &$0.306 - {\it i} 0.029$
     & $0.339 - {\it i} 0.057$
     &$0.94 \times 10^{-3}$
     & $1.16 \times 10^{-3}$
 \\
 \hline \hline
\end{tabular}}
\end{center}
\end{table}

At last, we calculate the $B_c \to J/\psi$ transition form factor $A_0^{B_c \to J/\psi}(0)$ and
the $B_c^+ \to J/\psi \pi^+$ branching ratio ${\rm BR}(B_c^+ \to J/\psi \pi^+)$ in the PQCD approach,
taking into account the NLL $k_T$ resummation effect from Eqs.~(\ref{sphm}) and (\ref{bm}).
The explicit expressions for the above quantities, together with the input parameters and the models
of the meson wave functions, can be found in \cite{Liu:2018kuo}. The initial
scale of the renormalization-group evolution for the meson wave functions,
governed by the quark anomalous dimension \cite{Liu:2018kuo},
is modified from $m_c$ to $m_c/\sqrt{2}$ for consistency. We adopt
the one-loop running formula for the strong coupling $\alpha_s$. It has been checked that the two-loop running
causes only 1-2\% reduction of the results from the one-loop running. The dependence of the
quantities $A_0^{B_c \to J/\psi}(0)$ and ${\rm BR}(B_c^+ \to J/\psi \pi^+)$ on the shape
parameter $\beta_{B_c}$ of the $B_c$ meson wave function in the range $[0.8, 1.2]$ GeV is presented in
Table~\ref{tab:d-beta-A0-Br}, and compared with that derived with the LL
resummation effect \cite{Liu:2018kuo}. The potential imaginary part of $A_0^{B_c \to J/\psi}(0)$,
which is supposed to be a real object \cite{Manohar:2000dt}, increases a bit under the NLL
resummation, but remains power suppressed. It is found that
$A_0^{B_c \to J/\psi}(0)$ is enhanced by the NLL resummation effect by 5-10\%, as $\beta_{B_c}$ varies
from 0.8 GeV to 1.2 GeV, and thus ${\rm BR}(B_c^+ \to J/\psi \pi^+)$ increases by
about 10-20\% accordingly. It implies that the NLL resummation effect is not negligible,
and crucial for the determination of the $B_c$ meson wave function, when relevant
data are available in the future. The values in Table~\ref{tab:d-beta-A0-Br}
are consistent with those from other approaches in the literature, which have been
summarized in \cite{Liu:2018kuo}.

\section{CONCLUSION}

In this letter we have improved the $k_T$ resummation for the $B_c\to J/\psi$ decays,
which involve an additional intermediate charm scale compared with the $B\to\pi$ decays,
to the NLL accuracy. We constructed the evolution equation for the
TMD meson wave function by varying its associated
gauge link, performed the $k_T$ resummation by solving the evolution equation, and fixed
the NLL pieces through the matching to the one-loop calculation. Our work represents the
first NLL $k_T$ resummation with the multiple scales $m_b$, $m_c$ and $k_T$
for $B_c$ meson decays. It has been observed that the NLL resummation effect enhances the
$B_c \to J/\psi$ transition form factor and the
$B_c^+ \to J/\psi \pi^+$ branching ratio more than the LL resummation effect does.
With more precise data from future experiments and the more accurate resummation formula
obtained here, it is possible to determine the shape parameter of the $B_c$ meson
wave function, which can then be adopted to make reliable predictions for other decay modes.
The above improved PQCD approach is also applicable to $B$ and $B_c$ meson decays to
charmonia. Based on this work, we are ready to extend the $k_T$ resummation with
multiple scales to energetic charmed mesons, for which the current formalism is still preliminary.

\begin{acknowledgments}

This work is supported in part by
the Ministry of Science and Technology of R.O.C. under
Grant No. MOST-107-2119-M-001-035-MY3, by the National Natural Science
Foundation of China under Grants No.~11875033 and No.~11775117, by the 
Qing Lan project of Jiangsu Province under Grant No.~9212218405, 
and by the Research Fund of
Jiangsu Normal University under Grant No.~HB2016004.
\end{acknowledgments}


\end{document}